\documentclass[a4paper,11pt]{article}
\usepackage{jheppub} 
\pdfoutput=1
\usepackage[utf8]{inputenc}
\usepackage{hyperref}
\usepackage{amsfonts}
\usepackage{amsmath}
\usepackage{amssymb}
\usepackage{mathrsfs}  
\usepackage{color}
\usepackage{physics}
\usepackage{multicol}
\usepackage{tikz}

\usepackage{amsmath,bbm,array,amsfonts,graphicx,wrapfig,lscape,float,mathtools,multirow,longtable,amsthm}
\usepackage{stackrel}
\usepackage[all]{xy}
\usepackage{caption}
\usepackage{subcaption}
\usepackage{multirow}
\usepackage[bottom]{footmisc}
\usepackage{mathrsfs}
\usepackage{tikz}
\usepackage{bm}
\usepackage{color}
\usepackage{enumerate}
\usetikzlibrary{decorations.pathreplacing}
\usepackage{diagbox}
\numberwithin{equation}{section}
\numberwithin{figure}{section}
\numberwithin{table}{section}
\usepackage{pgfplots}
\pgfplotsset{compat=1.14}
\usepackage{makecell}
\usepackage{lscape}
\usepackage[utf8]{inputenc}
\usepackage{mathtools}
\usepackage{todonotes}
\usepackage{epigraph}
\usepackage{boldline}

\begin{document}

\title{Machine Learning for Hilbert Series}
\author[a,b]{Edward Hirst}
\affiliation[a]{Department of Mathematics, City, University of London, EC1V 0HB, UK}
\affiliation[b]{London Institute for Mathematical Sciences, Royal Institution, London W1S 4BS, UK}
\emailAdd{edward.hirst@city.ac.uk}

\abstract{Hilbert series are a standard tool in algebraic geometry, and more recently are finding many uses in theoretical physics. This summary reviews work applying machine learning to databases of them; and was prepared for the proceedings of the Nankai Symposium on Mathematical Dialogues, 2021.
}

\maketitle

\section{Introduction}\label{sec:intro}
Mathematically, Hilbert series (HS) provide a means of counting regular functions on varieties. 
Physically, they are used in a range of scenarios, including to count BPS operators \cite{Benvenuti:2006qr,Feng:2007ur,Gray:2008yu,Hanany:2008sb,Chen:2011wn,Jokela:2011vg}, determine moduli spaces \cite{Benvenuti:2010pq,Hanany:2012dm,Buchbinder:2019eal,Forcella:2008ng}, find standard model invariants \cite{Hanany:2010vu,Lehman:2015coa,Xiao:2019uhh}, deal with string compactifications \cite{Braun:2012qc}, construct effective Lagrangians \cite{Lehman:2015via,Henning:2015daa,Kobach:2017xkw,Anisha:2019nzx,Marinissen:2020jmb,Graf:2020yxt}, as well as for a range of other quiver-related computations \cite{Cremonesi:2013lqa,Bourget:2020xdz,Bourget:2021siw,Bourget:2019rtl,Bourget:2021xex,Closset:2020scj,Franco:2005sm,Argurio:2020dko,Franco:2022ziy,Dey:2013fea,Flo2021}.

In parallel, machine learning (ML) is fast becoming a definitive methodology for data analysis and pattern recognition. 
The most common tool being neural networks (NNs), which are many-parameter non-linear functions used to learn properties of a dataset.
Beyond supervised methods, principal component analysis (PCA) is an established unsupervised method, used for examining clustering in datasets.
PCA puts focus on a data representation's largest principal components, which are found through diagonalisation of the data's covariance matrix.
ML has also been successfully applied to a variety of physically motivated scenarios, including: Calabi-Yau manifolds \cite{He:2017aed,He:2017set,Carifio:2017bov,Krefl:2017yox,Ruehle:2017mzq,Candelas:2016fdy,Ashmore:2019wzb,Berman:2021mcw,Bao:2020sqg}, polytopes \cite{Bao:2021ofk,Berglund:2021ztg}, graph theory \cite{He:2020fdg}, knot theory \cite{Craven:2021ckk,Gukov:2020qaj}, amoebae \cite{Bao:2021olg}, brane webs \cite{Arias-Tamargo:2022qgb}, integrability \cite{Krippendorf:2021lee}, Seiberg duality among quivers \cite{Bao:2020nbi}, and the related dessin d'enfant Galois orbits \cite{He:2020eva}.

In this summary, work from \cite{Bao:2021auj} using neural networks to learn geometric properties of a variety from coefficients in its respective HS is described. 
The work was carried out in \texttt{python} with the use of \texttt{Tensorflow} \cite{tensorflow2015-whitepaper} for machine learning implementation. 


\section{Defining Hilbert Series}\label{HS}

Within algebraic geometry HS encode information about a complex variety $X$, embedded in complex weighted projective space $\mathbb{P}_\mathbb{C}(p_0^{q_0}, \ldots, p_s^{q_s})$, such that weight $p_i$ occurs $q_i$ times. 
The variety's coordinate ring is denoted $R=\mathbb{C}[x_0,\ldots,x_k]/I$, for complex variables $x_i$, and homogeneous ideal $I$ generated by the polynomials defining the variety.
Any specific embedding of the variety then induces a grading on this coordinate ring \cite{Dolga}, which can be written $R=\bigoplus_{i\geq 0} R_i$.
The HS is the generating function for the dimensions of these graded pieces of $R$,
\begin{equation}\label{hs_eqn}
    H(t; X) = \sum\limits_{i=0}^{\infty} (\dim_{\mathbb{C}} R_i) t^i \;;
\end{equation}
such that $\dim_{\mathbb{C}}R_i$ defines the number of independent degree $i$ polynomials on the variety, and $t$ is a dummy variable used to keep track of the grading order (known as the fugacity in the context of quivers).

This concept can be generalised to any $n$-multiple grading of the ring, such that\\ $R=\bigoplus_{i_1,i_2,...,i_n \geq 0} R_{i_1,i_2,...,i_n}$, giving HS
\begin{equation}
    H(t_1,t_2,...,t_n; X) = \sum\limits_{i_1,i_2,...,i_n=0}^{\infty} (\dim_{\mathbb{C}} R_{i_1,i_2,...,i_n}) t_1^{i_1}t_2^{i_2}...t_n^{i_n} \;;
\end{equation}
which in the physics literature is called the `refined' HS and each fugacity is a gauge invariant operator. Note it is simple to unrefine by setting $t_k=t \ \forall \ k$.

\begin{wrapfigure}[12]{r}{0.4\textwidth}
    \vspace{-20pt}
    \begin{center}
    \includegraphics[width=45mm]{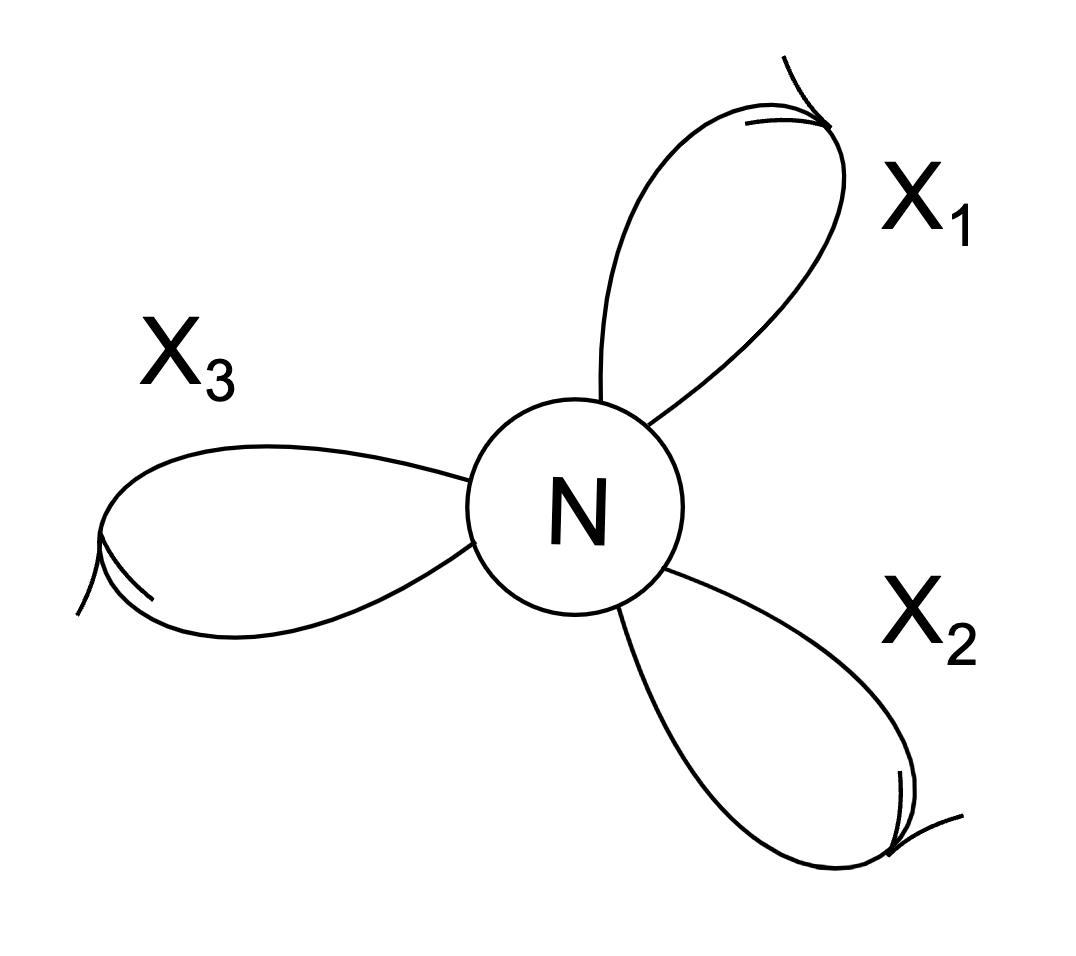}
    \caption{The quiver for a $\mathcal{N}=1$ $U(N)$ gauge theory with three adjoint chiral multiplets.}
    \label{quiver}
    \end{center}
\end{wrapfigure}

In the context of supersymmetric gauge theory in string theory, when one knows the matter content and superpotential defining the theory they often wish to compute the geometric information about the moduli space of vacua.
Whilst in the case of the moduli space being toric one can determine it from the combinatorics of the toric diagram \cite{Hanany:2005ss}, where the moduli space is not toric the process is more difficult. 
Hence in general, single-trace gauge invariant operators must be counted order by order to construct the HS to then learn about this moduli space and the underlying variety it is a cone over.

The process of computing a HS is nicely exemplified with the prototypical example of the clover quiver in $\mathcal{N}=1$ supersymmetry, see figure \ref{quiver}, with $U(N)$ gauge group and superpotential given by tracing over a certain choice of cycle combinations: $W=\Tr(X_1X_2X_3 - X_1X_3X_2)$.
The $X_k$ are the chiral supermultiplets in a representation (here adjoint) of the gauge group, and can be related to the fugacities $t_k=\Tr(X_k)$ in this case as $\Tr(X_k)$ for $k \in \{1,2,3\}$ represent the 3 order 1 generators.

Higher order generators are found using the information of the superpotential choice, computing all the F-terms as $0=\partial_{X_1}W=\partial_{X_2}W=\partial_{X_3}W$ shows all the 3 generators are freely commuting such that single trace operators can be constructed at each order from combining the operators in all ways.

Therefore in the refined case we have $\begin{psmallmatrix}n+2\\2\end{psmallmatrix}=\frac{(n+1)(n+2)}{2}$ single-trace operators at each order $n$, of the form $\Tr(X_1^aX_2^bX_3^c)$ such that $(a,b,c)$ are all possible combinations that sum to $n$. 
Hence the HS starts
\begin{alignat*}{3}
    \text{Order} & \ | \ \# \ && | \ \text{Operators}\\
    0 & \ | \ \, 1 \ && | \ 1 \\
    1 & \ | \ \, 3 \ && | \ \text{Tr}(X_1), \ \text{Tr}(X_2), \ \text{Tr}(X_3) \\
    2 & \ | \ \, 6 \ && | \ \text{Tr}(X_1^2), \ \text{Tr}(X_2^2), \ \text{Tr}(X_3^2), \ \text{Tr}(X_1X_2), \ \text{Tr}(X_1X_3), \ \text{Tr}(X_2X_3)
\end{alignat*}
for the first 3 orders. Consequently, if we consider the simplest case of $N=1$ where the multiplet representations, $X_k$, are just complex numbers, the HS in the refined and unrefined cases are
\begin{align}
    H(t_1,t_2,t_3;X) & = \frac{1}{(1-t_1)(1-t_2)(1-t_3)}\;,\\
    H(t;X) & = \frac{1}{(1-t)^3}\;,\label{example_hs2}
\end{align}
respectively. Since there are 3 independent generators with no relations (seen via the trivial numerator), the moduli space is therefore simply $\mathbb{C}^3$.
This can be shown by taking the plethystic logarithm of the HS where the first positive terms in the expansion are the generators, and the next are the relations (of which here there are none).

Note one may consider higher $N$ where the moduli space is then the $N$-th symmetric product of $\mathbb{C}^3$, deriving from the permutations of the $X_k$ matrices being invariant under the trace.
Additionally, using the plethystic exponential creates multi-trace operators of the form $\Tr(...)\Tr(...)...\Tr(...)$, which are also valid gauge-invariant operators but are instead from analysis of these symmetric products of the moduli space \cite{Benvenuti:2006qr}.

In the work summarised here the focus is a single grading, and hence the unrefined HS, as in equation \eqref{hs_eqn}, which via Hilbert-Serre theorem \cite{AtiyahMacdonald} can be written in two equivalent forms:
\begin{align}
    H(t; X) & = \frac{P(t)}{\prod\limits_{i=0}^s(1-t^{p_i})^{q_i}}\;, \label{hs_form1}\\
    H(t; X) & = \frac{\tilde{P}(t)}{(1-t^j)^{dim+1}}\;,\label{hs_form2}
\end{align}
for polynomials $P$ and $\tilde{P}$ in the fugacity $t$, and where the variety, $X$, has Gorenstein index $j$ and dimension $dim$.
From \eqref{hs_form2} one can tell if the variety is Gorenstein if and only if the numerator polynomial $\tilde{P}$ is palindromic \cite{Stanley1978}. 
Likewise one can also use the HS to distinguish whether a variety is a complete intersection, which happens when the expansion of the HS plethystic logarithm terminates at finite order \cite{Benvenuti:2006qr,Feng:2007ur}. 

We can therefore see that our example HS in \eqref{example_hs2} is trivially Gorenstein, and also trivially a complete intersection as their are no relations (and hence also no syzygies) among the generators to occur in the plethystic logarithm expansion.
One can also extract the dimension, however since the moduli space is formed from taking a specific cone over the variety its dimension is one larger than that given by equating \eqref{example_hs2} to \eqref{hs_form2}.
For our example it hence has a 3-dimensional $\mathbb{C}^3$ moduli space, which may be considered to be a cone over the 2-dimensional Fano $\mathbb{P}_{\mathbb{C}}^2$ variety.

\section{Machine Learning}
With data about the low order terms in the HS, one wishes to determine geometric information about the underlying algebraic variety.
In the work summarised here, NN regressors were used to predict the defining weights $(p_i)$ of the weighted projective space each variety is embedded in, and NN classifiers learnt the Gorenstein index $j$ and dimension $dim$ of the variety itself.
Further to these, three binary classification investigations were performed with NN classifiers to determine whether input HS expansion coefficients came from varieties which were Gorenstein, were complete intersections, or were `fake' such that they were generated from sampling as described below.

\paragraph{Data Information}
HS data used for this work was retrieved from the~GRDB~\cite{grdb,fanodata,kasprzyk2010canonical}, consisting of HS associated to three-dimensional~$\mathbb{Q}$-Fano varieties with Fano index one \cite{ABR:Fano,BK22}.
Investigations also used generated `fake' data, created through sampling the hyperparameters of the HS equations \eqref{hs_form1} and \eqref{hs_form2} from distributions fitted to the `real' GRDB data.
The sampling discarded hyperparameter combinations which did not satisfy certain physical conditions, encouraging the data to be more representative; and also discarded any repeats of GRDB data.

For both HS styles of input data mentioned, which we call `real' and `fake' respectively, the HS closed forms were Taylor expanded and expansion coefficients saved. 
These lists of coefficients form vectors of integers, and make up the ML input for the investigations carried out.
Two types of coefficient vector were used as inputs, one from the start of the series (coefficients 0-100) and one from deeper into it (coefficients 1000-1009).
The reason to also learn with coefficients sampled from deeper in the series was to provide intuition on the importance of the variety's orbifold points, which have a greater relative significance than the `initial part' at higher orders.

Output data depended on the investigation.
For the regression of embedding weights, these were sorted 3-vectors of integers (sampled in the range [1,10]).
For the multi-classification the outputs were single integers in the range [1,5], which were the values of either the variety's Gorenstein index or its dimension respectively.
The final binary classifications just outputted 0 or 1 dependent on whether the Gorenstein or complete intersection properties were respectively satisfied, or whether the data came from the GRDB or was generated by us.

\paragraph{NN Architectures} 
For all investigations the NNs used were built of 4 dense layers of 1024 neurons, all with ReLU activation and 0.05 dropout factor. 
Training was in batches of 32 for 20 epochs using the Adam \cite{Adam} optimiser to minimise either the regression log(cosh) loss function or the classification cross-entropy loss function.
5-fold cross-validation provided confidence on the metrics used to evaluate the learning via averaging and standard error. 

In the regression investigation, mean absolute error (MAE) evaluated learning by computing the average absolute difference between each true embedding weight and the NN predicted embedding weight.
This measure evaluates in the range $[0,\infty)$ with 0 indicating perfect learning, where the embedding weights are always correctly predicted.
In the classification investigations, Matthew's correlation coefficient (MCC), which is an application of Pearson's correlation coefficient to binary variables, evaluated the learning.
This measure evaluates in the range $[-1,1]$ with 1 indicating perfect learning.

\paragraph{Results}
ML results are provided in table \ref{ml_results} for each of these 5 investigations.
Learning of the varieties' ambient space embedding weights predicted each weight within an average range of 1 from the correct values when using higher order coefficients.
Additionally the geometric parameters of equation \eqref{hs_form2} were learnt exceptionally well, but now better from the lower order coefficients.

Each of the binary classifications also evaluated with MCC values exceeding 0.9, demonstrating strong learning. 
Both the Gorenstein and `fake' (non-GRDB) properties were better identified with higher order coefficients, whilst the complete intersection property was only learnt from lower orders but performed equally well.

All these results demonstrate the success of ML in learning geometric properties of varieties from HS, and hopefully further encourages the use of ML techniques in the context of Hilbert series and beyond.

\begin{table}[!tb]
\centering
\scriptsize
\begin{tabular}{|c|c|c|c|c|c|c|c|} 
\hline
\multicolumn{2}{|c|}{\begin{tabular}[c]{@{}c@{}}\\Investigation\end{tabular}} & \begin{tabular}[c]{@{}c@{}}Embedding\\ weights\end{tabular}                & \begin{tabular}[c]{@{}c@{}}Gorenstein\\ index\end{tabular}                   & Dimension                                                                    & \begin{tabular}[c]{@{}c@{}}Gorenstein\\ property\end{tabular}                & \begin{tabular}[c]{@{}c@{}}Complete\\ Intersection\end{tabular}              & GRDB \\ 
\hline
\multicolumn{2}{|c|}{Output Ranges}                                           & 3 x [1,10] & [1, 5] & [1, 5] & Binary & Binary  & Binary \\ 
\hline
\multicolumn{2}{|c|}{Measure}                                                 & MAE & MCC & MCC & MCC & MCC & MCC \\ 
\hline
\multirow{2}{*}{Orders} & 0 - 100                                             & \begin{tabular}[c]{@{}c@{}}1.94 \\$\pm$ 0.11\end{tabular} & \begin{tabular}[c]{@{}c@{}}0.916 \\$\pm$ 0.010\end{tabular} & \begin{tabular}[c]{@{}c@{}}0.993 \\$\pm$ 0.006\end{tabular} & \begin{tabular}[c]{@{}c@{}}0.717 \\$\pm$ 0.155\end{tabular} & \begin{tabular}[c]{@{}c@{}}0.910 \\$\pm$ 0.022\end{tabular} & \begin{tabular}[c]{@{}c@{}}0.717\\$\pm$ 0.155\end{tabular}  \\ 
\cline{2-8}
& 1000 - 1009 & \begin{tabular}[c]{@{}c@{}}1.04 \\$\pm$ 0.12\end{tabular} & \begin{tabular}[c]{@{}c@{}}0.727 \\$\pm$ 0.022\end{tabular} & \begin{tabular}[c]{@{}c@{}}0.822 \\$\pm$ 0.031\end{tabular} & \begin{tabular}[c]{@{}c@{}}0.919 \\$\pm$ 0.073\end{tabular} & - & \begin{tabular}[c]{@{}c@{}}0.919\\$\pm$ 0.073\end{tabular}  \\
\hline
\end{tabular}
\caption{Machine learning results for each of the investigations. Neural networks learnt geometric properties associated to an algebraic variety from coefficients in their Hilbert series expansions. Regression was assessed with mean absolute error (MAE) and classification with Matthew's correlation coefficient (MCC), training using 5-fold cross-validation to provide a standard error on the averaged measures.}\label{ml_results}
\end{table}

\paragraph{PCA}
Whilst all the learning so far uses NNs from supervised ML, PCA provides another avenue for analysis of this data from the conjugate field of unsupervised ML.

\begin{figure}[!tb]
	\centering
	\begin{subfigure}{0.32\textwidth}
    	\centering
    	\includegraphics[width=0.95\textwidth]{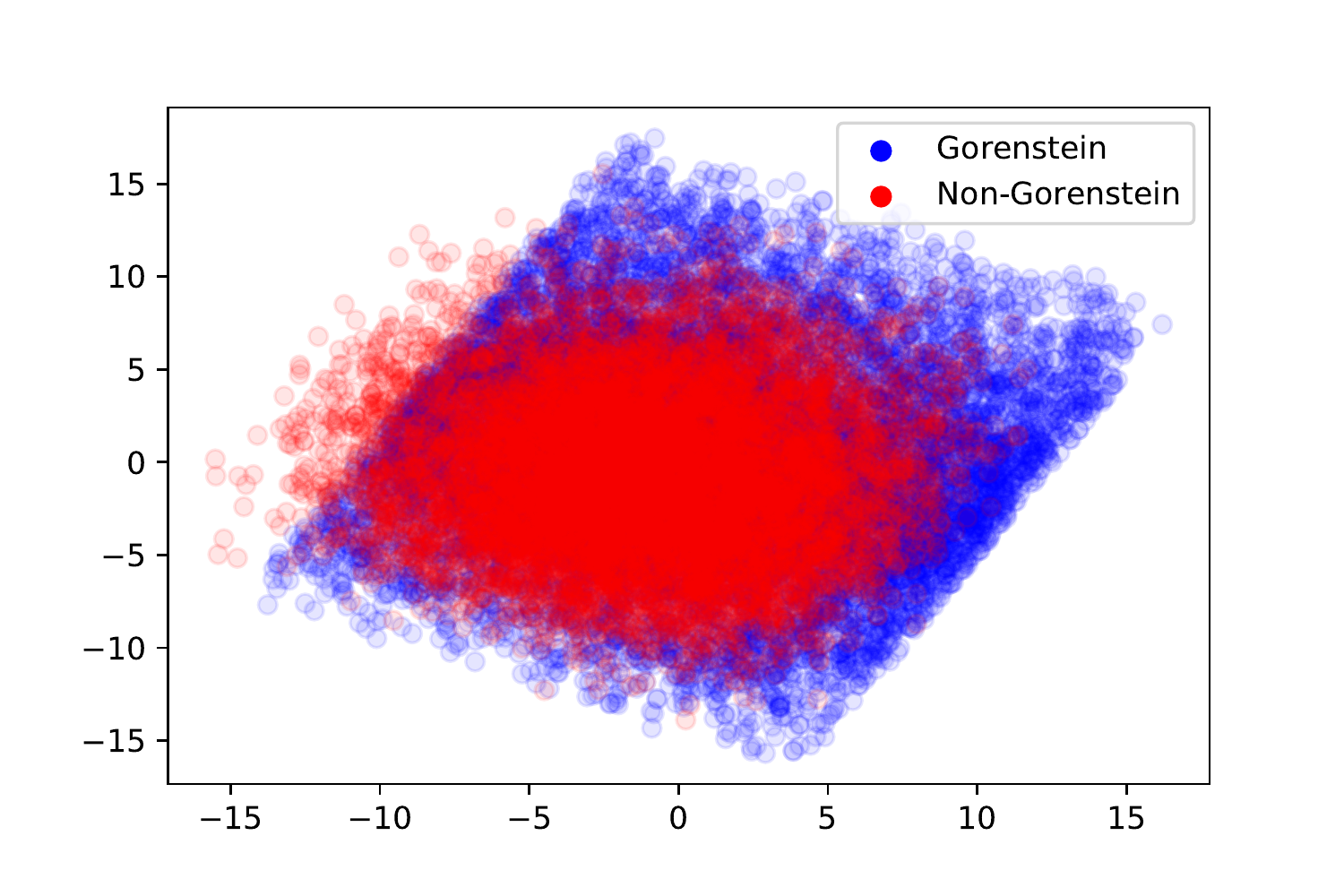}
    	\caption{Gorenstein}\label{j_PCA}
	\end{subfigure} 
    \begin{subfigure}{0.32\textwidth}
    	\centering
    	\includegraphics[width=0.95\textwidth]{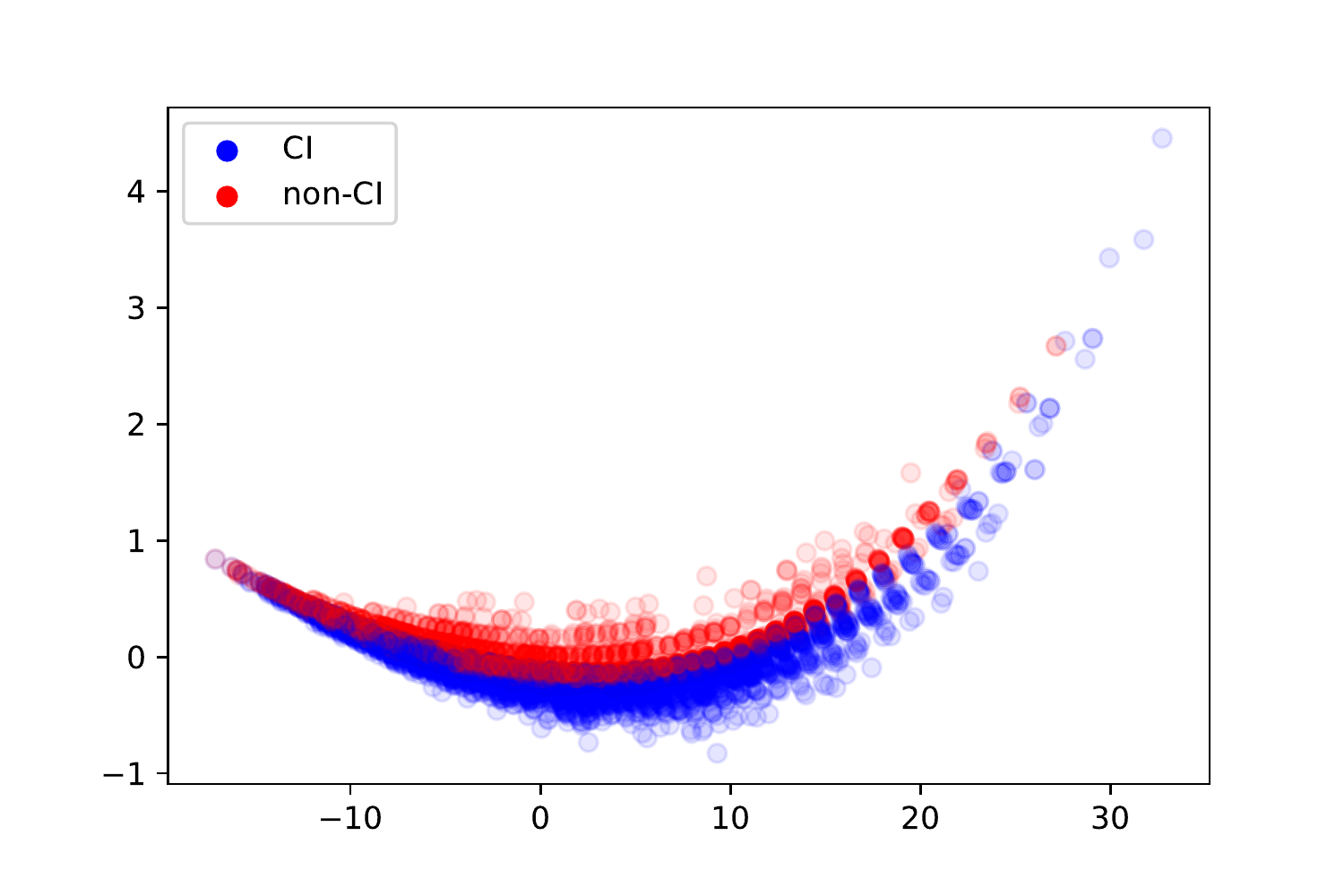}
    	\caption{Complete Intersection}\label{CI_PCA}
    \end{subfigure} 
    \begin{subfigure}{0.32\textwidth}
    	\centering
    	\includegraphics[width=0.95\textwidth]{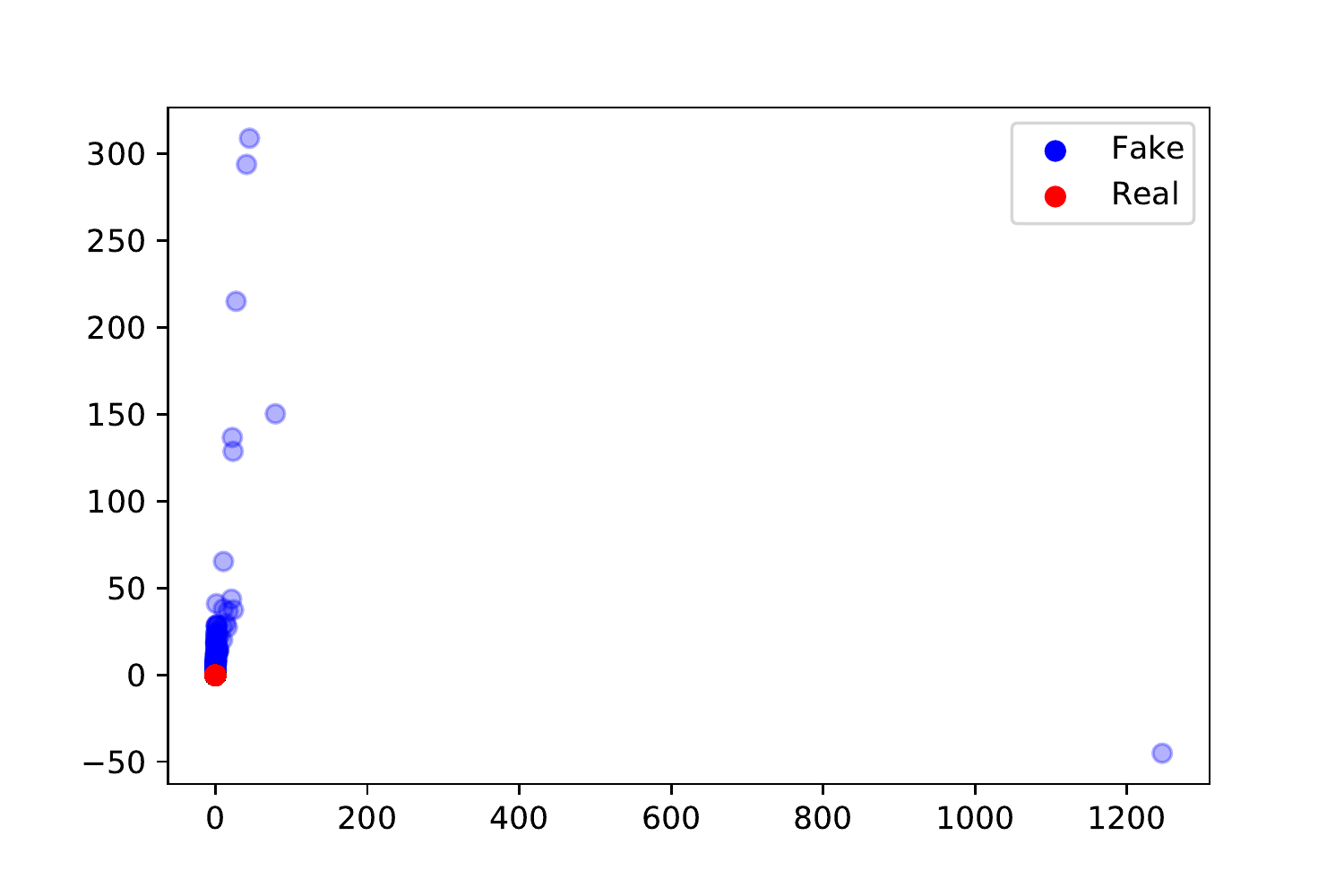}
    	\caption{`real' / `fake'}\label{RF_PCA}
    \end{subfigure}
\caption{2d principal component analysis plots of the Hilbert series 0-100 coefficient data, labelled by each of the binary classification investigations respectively.}\label{PCA}
\end{figure}

In PCA, the covariance matrix is computed between all vector entries, which is then subsequently diagonalised and eigenvectors sorted in decreasing order of variance eigenvalue. 
Each data vector is then re-expressed in this eigenvector basis, which in this case of considering the order 0-100 coefficients gives vectors with 101 entries.
Since this is too high-dimensional to visualise, the first 2 principal components of each datapoint are plotted to provide a means for examining the data's clustering, these plots can be seen for the 3 binary classification investigations in figure \ref{PCA}. 

Here, 2d PCA was performed for each of the binary classification investigations, and the datapoints in each class plotted separately in different colours.
Whilst there is no clear separation for the Gorenstein data, the complete intersection and `real'/`fake' data do show some separation, indicating there is some linear structure that the NNs can take advantage of for classification. 
In particular the occurrence of this for the `fake' data demonstrates the difficulty of generating representative HS.

\section*{Acknowledgements}
The author wishes to thank Jiakang Bao, Yang-Hui He, Johannes Hofscheier, Alexander Kasprzyk, and Suvajit Majumder for collaboration on this work; Jiakang Bao again for clarifying discussion; and also STFC for the PhD studentship.

\linespread{0.9}\selectfont
\addcontentsline{toc}{section}{References}
\bibliographystyle{utphys}
\bibliography{references}

\end{document}